\documentclass[a4paper]{article}
\usepackage{amssymb}
\usepackage{amsmath}
\usepackage{bm}
\usepackage{multirow}
\usepackage{mathtools}
\usepackage{geometry}
\usepackage[toc]{appendix}
\usepackage{longtable}

\title{Linkage Free Dual System Estimation}
\author{Viktor Ra\v{c}inskij\footnote{vr1v14@soton.ac.uk} $^{1, \textrm{ } 2}$, Paul A. Smith\footnote{p.a.smith@soton.ac.uk} $^1$, Peter G. M. van der Heijden\footnote{p.g.m.vanderheijden@uu.nl} $^{1, \textrm{ } 3}$}

\date{%
	$^1$Department of Social Statistics and Demography, University of Southampton\\%
	$^2$Office for National Statistics\footnote{All views expressed are those of the author and do not necessarily reflect the views of ONS}\\
	$^3$Department of Social Sciences, Methodology and Statistics, Utrecht University\\
	\bigbreak
	First version\\ 26$^\textrm{th}$ March, 2019
}

\begin{document}

\maketitle

\begin{abstract}
In this paper it is shown that under certain conditions there is a relationship between the parameter estimation of the Fellegi--Sunter probabilistic linkage model and dual system estimation. This relationship can be used as the basis of an approach to population size estimation. In this case it is sufficient to estimate the parameters of the linkage model in order to obtain the population size estimate. Neither classification of the record pairs into links and non-links, nor forcing the records into a series of 1-1 matches, nor clerical review of the potential links is required. The accuracy of the proposed estimator appears to be bounded by the accuracy of the dual system estimator with perfect linkage and it diminishes as the discriminatory power of the linkage variables decreases.
\end{abstract}


\section{Introduction}
Capture-recapture methods provide a way to estimate the size of a population from multiple but incomplete surveys of that population. These methods have been used to estimate the size of animal populations \cite{seber82, mccreamor15}, the under-enumeration of a human population census \cite{wolter86, brownetal18}, the number of war casualties \cite{balletal02}, the number of duplicates in a database \cite{herzogetal07}, the extent of human trafficking \cite{silverman18}, the number of homeless people \cite{coumansetal17} and many more similar examples. 

To enable the capture-recapture estimation, the surveys (or lists) are linked together. Usually, the linkage process must be perfect to guarantee that the capture-recapture estimator is unbiased. Thus, capture-recapture methods are closely related to the fields of record linkage \cite{fellegisunter69, herzogetal07, christen12} and estimation of the linkage error \cite{biemer88, tuoto16}. The linkage error adjusted population size estimation \cite{dingfienb94, dewolfetal18} combines together capture-recapture, record linage and estimation of the linkage error. In order to correct for the linkage errors in capture-recapture estimation, existing methods require an estimate of the linkage errors that occurred at the linkage stage. The estimation of the error rates often depends on a supplementary data collection exercise \cite{dingfienb94}.

In this paper a new method to estimate the population size from two incomplete lists of the population elements is proposed (i.e., we are dealing with a special case of capture-recapture known as dual system estimation). This method requires only estimation of the linkage model parameters and does not require the record pairs to be classified as links and non-links. Neither does it require additional data collection as the basis for estimation of the linkage errors. Hence, various stages of the record linkage process (such as setting the linkage thresholds, clerical resolution of the potential links, and forcing the records into a series of 1-1 matches) and complexities of the linkage error adjusted population size estimation are avoided. Furthermore, the cost of data processing and estimation is expected to be reduced since there is no need to review the unclassified record pairs clerically. Therefore, the proposed estimator offers a number of advantages over the simple dual system estimator (DSE) or existing linkage error corrected DSEs. All the assumptions of the DSE except the assumption of perfect linkage must hold. The assumptions of the Fellegi--Sunter probabilistic linkage model \cite{fellegisunter69} must also hold. The proposed method has lower accuracy compared to the simple DSE when there is perfect linkage. However, in presence of the linkage errors, as the net linkage error rate increases, the proposed method becomes more accurate than the DSE.

\section{Dual system estimator}
\label{dse}
The DSE is one of the many capture-recapture estimators \cite{seber82, mccreamor15} for the population size $N$. The distinctive feature of this estimator among the two lists (or two capture) focused estimators is that it allows for variation in the list inclusion (or capture) probabilities \cite{wolter86}. Every element $i$ in a population of interest $U$ of size $N$ has probability $p_{i1+} = p_{1+}$ to be included in the first list and the probability $p_{i+1} = p_{+1}$ to be included in the second list, and in general $p_{1+} \neq p_{+1}$. Note, however, that the equality of individual inclusion probabilities in each list may be relaxed: it is sufficient to have equal inclusion probabilities on one of the sources or have uncorrelated inclusion probabilities \cite{seber82}.

We denote the number of population elements included in the first list by $n_{1+}$,  the number of population elements included in the second list by $n_{+1}$, and the number of population elements included in both list by $n_{11}$. The likelihood function for the DSE model is given by
\[L\left(N, p_{1+}, p_{+1}\right) = {{N}\choose{n_{11}, n_{1+}, n_{+1}}}p_{1+} ^ {n_{1+}}p_{+1} ^ {n_{+1}}(1 - p_{1+}) ^ {N - n_{1+}}(1 - p_{+1}) ^ {N - n_{+1}},\]
where the first term on the right hand side is the multinomial coefficient.

The maximum likelihood estimator of $N$ is then given by
\[\hat{N} = \left\lfloor{\frac{n_{1+}n_{+1}}{n_{11}}}\right\rfloor,\]
where $\lfloor\cdot\rfloor$ is the greatest integer function \cite{pollock76, wolter86}. However, in practice it is common to simply use the following estimator
\begin{equation}
\hat{N} = \frac{n_{1+}n_{+1}}{n_{11}}.
\end{equation}

Besides the constant inclusion probability of elements in at least one of the lists (or zero correlation between the two lists' inclusion probabilities) the DSE requires a number of additional assumptions to hold: a closed population, causal independence between the two lists, absence of spurious events (duplication, over-count, etc.) and perfect linkage (matching) \cite{wolter86}. 

Before the linkage takes place, all we observe are two individual lists and hence each list's population counts: $n_{1+}$ and $n_{+1}$ (the marginal counts). The record linkage procedure is required to establish $n_{11}$. The perfect linkage assumption means that $n_{11}$ can be established without error.

\section{Record linkage}
\label{rl}
Record linkage may be defined as a process of establishing whether two or more records represent the same element in a population when unique element identifiers are not available \cite{herzogetal07, christen12}. In this paper we focus on the probabilistic linkage model proposed by Fellegi and Sunter \cite{fellegisunter69}. Below is a brief summary of this model.

We consider linking two data sets $L_{1}$ and $L_{2}$ that have a certain number of elements in common. The set of all ordered pairs is $\Omega = L_{1} \times L_{2} = \{(l_{1}, l_{2}): l_{1} \in L_{1}, l_{2} \in L_{2} \}$. The set $\Omega$ can be partitioned into two disjoint sets. The first one, $M = \{(l_{1}, l_{2}): l_{1} = l_{2}, l_{1} \in L_{1}, l_{2} \in L_{2} \}$, is called the \textit{matched set} (or simply the \textit{m-set}) and consists of all pairs of records that represent the same elements in a population. The second one, $U = \{(l_{1}, l_{2}): l_{1} \ne l_{2}, l_{1} \in L_{1}, l_{2} \in L_{2} \}$, is called the \textit{unmatched set} (\textit{u-set}) and consists of all pairs of records that do not represent the same elements in the population.

Every record has a set of variables (such as first name, surname, age, etc.) associated with it. These variables are labelled $v = 1, \dots, k$. For some of the records certain variables may have missing values. For every pair of records, $j = 1, \dots, |\Omega|$, where $|\cdot|$ denotes the size of a set, a comparison vector, $\boldsymbol{\gamma} ^ j = (\gamma_{1} ^ j, \dots, \gamma_{k} ^ j)$, can be constructed. The components, $\gamma_{v} ^ j$, are outcomes of comparing the values of a variable $v$ in a record pair $j$. For simplicity and in order to be consistent with the development in \cite{jaro89} we restrict our discussion to $\gamma_{v} ^ j = \{0, 1\}$.

For each $\boldsymbol{\gamma} \in \Gamma$ ($\Gamma$ is the space of all possible comparison vectors) a linkage rule $d(\boldsymbol{\gamma})$ assigns probabilities of taking three possible linkage decisions (link, possible link, non-link), that is 
\[d(\boldsymbol{\gamma}) = \{\mathbb{P}(\text{link}\mid\boldsymbol{\gamma}), \mathbb{P}(\text{possible link}\mid\boldsymbol{\gamma}), \mathbb{P}(\text{non-link}\mid\boldsymbol{\gamma})\},\]
\[\sum{\mathbb{P}(\cdot\mid\boldsymbol{\gamma})} = 1.\]
Note that we use the terms \textit{match} and \textit{non-match} to denote the true status of a record pair, whereas we use the terms \textit{link} and \textit{non-link} to denote the inferred status of a record pair \cite{larsenrubin01}.

In order to formalise the errors associated with the linkage rule, we define the following conditional probabilities: $m(\boldsymbol{\gamma}) = \mathbb{P}(\boldsymbol{\gamma}\mid M)$ (\textit{m-probability}) and $u(\boldsymbol{\gamma}) = \mathbb{P}(\boldsymbol{\gamma}\mid U)$ (\textit{u-probability}). There are two errors related to the linkage rule $d(\boldsymbol{\gamma})$. The first one is the false positive (FP) error which occurs when a pair from the non-matched set is classified as a link. The corresponding probability is
\[\mu = \mathbb{P}(\text{link}\mid U) = \sum_{\gamma \in \Gamma}{u(\boldsymbol{\gamma})\mathbb{P}(\text{link}\mid \boldsymbol{\gamma})}.\]
The second error is the false negative (FN) which occurs when a pair from the match set is classified as a non-link. The corresponding probability is
\[\lambda = \mathbb{P}(\text{non-link}\mid M) = \sum_{\gamma \in \Gamma}{m(\boldsymbol{\gamma})\mathbb{P}(\text{non-link}\mid \boldsymbol{\gamma})}.\]

We say that the linkage rule $L(\mu, \lambda, \Gamma)$ is at the levels $\mu, \lambda$. The optimal rule $L(\mu, \lambda, \Gamma)$ is such that $\mathbb{P}(\text{possible link}\mid L) \le \mathbb{P}(\text{possible link}\mid L ^ {\prime})$ for every $L ^ {\prime}(\mu, \lambda, \Gamma)$.

Fellegi and Sunter \cite{fellegisunter69} demonstrated that for an admissible pair of error levels $(\mu, \lambda)$ the optimal linkage rule is given by
\[d(\boldsymbol{\gamma}) = 
\begin{cases} (1, 0, 0) \: \text{if} \: T_{\mu} \le m(\boldsymbol{\gamma}) / u(\boldsymbol{\gamma}) \\
(0, 1, 0) \: \text{if} \: T_{\lambda} < m(\boldsymbol{\gamma}) / u(\boldsymbol{\gamma}) < T_{\mu} \\
(0, 0, 1) \: \text{if} \: m(\boldsymbol{\gamma}) / u(\boldsymbol{\gamma}) \le T_{\lambda}
\end{cases} \]
where $T_{\mu} = m(\boldsymbol{\gamma}^{n}) / u(\boldsymbol{\gamma}^{n})$, $T_{\lambda} = m(\boldsymbol{\gamma}^{n ^ {\prime}}) / u(\boldsymbol{\gamma}^{n ^ {\prime}})$ are decision thresholds. In the above expression $n < n ^ {\prime}$ and are such that
\[\sum_{j = 1} ^ {n - 1}{u_{j}} < \mu \le \sum_{j = 1} ^ {n}{u_{j}} \: \text{and} \: \sum_{j = n ^ {\prime}} ^ {N_{\Gamma}}{m_{j}} \ge \lambda > \sum_{j = n ^ {\prime} + 1} ^ {N_{\Gamma}}{m_{j}} \] 
and $N_{\Gamma}$ is the number of unique comparison vectors. The interested reader is referred to the original paper \cite{fellegisunter69} to see how the ordering of $j$ is obtained.

\section{Parameter estimation for the linkage model}
\label{parest}
There are several ways of implementing the parameter estimation for the record linkage model described above. In this paper we focus on the maximum likelihood estimation method utilising the Expectation-Maximization (EM) algorithm \cite{dempsteretal77}. The method was proposed by Jaro \cite{jaro89} and it is a common alternative to the original Fellegi--Sunter method. In the summary below we follow \cite{jaro89, herzogetal07}.

Let $S = L_{1, s} \times L_{2, s} = \{(l_{1}, l_{2}): l_{1} \in L_{1, s}, l_{2} \in L_{2, s} \}$, where $L_{1, s}$ and $L_{2, s}$ are simple random samples from $L_{1}$ and $L_{2}$, respectively. The size of $S$ is $|S|$, with members $r_{j} \in S, j = 1, \dots, |S|$. We assume that the comparison vector $\boldsymbol{\gamma}$ follows a finite mixture distribution

\begin{equation}
\Phi(\boldsymbol{m}, \boldsymbol{u}, p) = \mathbb{P}(\boldsymbol{\gamma};\boldsymbol{m}, \boldsymbol{u}, p) = p\cdot\mathbb{P}(\boldsymbol{\gamma}\mid M) + (1 - p)\cdot\mathbb{P}(\boldsymbol{\gamma}\mid U)
\end{equation}
with unknown parameters, but with the known number (two) of components in the mixture. In this case, $\boldsymbol{m} = (m_{1}, \dots, m_{k})$ and $\boldsymbol{u} = (u_{1}, \dots, u_{k})$ are vectors of m- and u-probabilities associated with each comparison variable, $m_{v} = \mathbb{P}(\gamma_v\mid M)$, $u_{v} = \mathbb{P}(\gamma_v\mid U)$. The third parameter in the mixture, $p$, is the proportion of the matched records among $|S|$ record pairs, $p = |M|/|S|$.

The complete data vector is $(\boldsymbol{g}, \boldsymbol{\gamma})$,  where $\boldsymbol{g} = (g_{1}, \dots, g_{|S|})$ is a vector of components defined as
\[g_{j} = \begin{cases}
1\:\text{if}\: r_{j} \in M \\
0\:\text{if}\: r_{j} \in U.
\end{cases}\]

The complete data likelihood associated with the mixture is given by
\begin{equation}
f(\boldsymbol{g}, \boldsymbol{\gamma}\mid \boldsymbol{m}, \boldsymbol{u}, p) = \prod_{j = 1} ^ {|S|}[p\cdot\mathbb{P}(\gamma ^ j\mid M)] ^ {g_{j}}[(1 - p)\cdot\mathbb{P}(\gamma ^ j\mid U)] ^ {1 - g_{j}}.
\end{equation}

Assuming conditional independence of m- and u-probabilities for each comparison variable, that is
\begin{equation}
\mathbb{P}(\gamma ^ {j}\mid M) = \prod_{v = 1} ^ {k}{m_{v} ^ {\gamma_{v} ^ {j}}}(1 - m_{v}) ^ {1 - \gamma_{v} ^ {j}}
\end{equation}
and
\begin{equation}
\mathbb{P}(\gamma ^ {j}\mid U) = \prod_{v = 1} ^ {k}{u_{v} ^ {\gamma_{v} ^ {j}}}(1 - u_{v}) ^ {1 - \gamma_{v} ^ {j}},
\end{equation}
the EM algorithm can be used to find the solutions for $(\boldsymbol{m}, \boldsymbol{u}, p)$.

At the expectation step, we compute the expectation of the indicator function $g_j$
\[\hat{g}_j = \frac{\hat{p}\prod_{v = 1} ^ {k} {\hat{m}_{v} ^ {\gamma_{v} ^ {j} } (1 - \hat{m}_{v}) ^ {1 - {\gamma_{v} ^ {j}} } }}{\hat{p}\prod_{v = 1} ^ {k} {\hat{m}_{v} ^ {\gamma_{v} ^ {j} } (1 - \hat{m}_{v}) ^ {1 - {\gamma_{v} ^ {j}} } } + (1 - \hat{p})\prod_{v = 1} ^ {k} {\hat{u}_{v} ^ {\gamma_{v} ^ {j} } (1 - \hat{u}_{v}) ^ {1 - {\gamma_{v} ^ {j}} } }}. \]

At the maximization step, we obtain $\hat{m}_v, \hat{u}_v$ and $\hat{p}$, namely

\[\hat{m}_{v} = \frac{\sum_{j = 1} ^ {|S|}{\hat{g}_{j}\gamma_{v} ^ {j} } }{\sum_{j = 1} ^ {|S|}{\hat{g}_{j}} },\]

\[\hat{u}_{v} = \frac{\sum_{j = 1} ^ {|S|}{(1 - \hat{g}_{j})\gamma_{v} ^ {j} } }{\sum_{j = 1} ^ {|S|}{(1 - \hat{g}_{j})} }\]
and
\[\hat{p} = \frac{\sum_{j = 1} ^ {|S|}{\hat{g}_{j}}}{|S|}. \]

At the first step of the algorithm we use some initial guess for the parameters and at iteration $\iota$ we use the estimates obtained at iteration $\iota - 1$.

\section{Linkage free dual system estimator}
\label{lfdse}
Probabilistic linkage involves an inevitable trade-off between the level of admissible errors (and hence the quality of estimation) and the amount of clerical resolution needed to resolve the set of potential links (and hence the time and cost of processing). The lower the level of admissible errors, the more clerical resolution is needed. In addition, many applications (including the DSE) require the so called 1-1 match constraint: every record on list $L_{1}$ ($L_{2}$) is either linked to \textit{one and only one} record on list $L_{2}$ ($L_{1}$) or unlinked. 

In this section we propose an approach which, under a certain probabilistic linkage set-up, allows the population size to be estimated using only the estimates of the probabilistic linkage parameters $(\hat{\boldsymbol{m}}, \hat{\boldsymbol{u}}, \hat{p})$. Hence, there is no need for clerical resolution or 1-1 constraining. Effectively, there is no linkage process as such, just the estimation of the linkage model parameters. The accuracy of the proposed estimator approaches the accuracy of the DSE with perfect linkage as the discriminatory power of linkage variables increases. When linkage error is present in the DSE, the proposed estimator may perform similarly or better depending on a combination of input parameters. Hereafter, we refer to our approach as the linkage free dual system estimation (LFDSE). Note that in contrast to linkage error corrected DSE methods \cite{dingfienb94, dewolfetal18}, which need estimates of the linkage error rates, the LFDSE does not require any additional data.

In what follows, we take it that all assumptions of the linkage model described above as well as all assumptions of the DSE, except the perfect linkage one, hold. 

Let an estimation stratum to be equivalent to a probabilistic linkage block. That is, the linkage and estimation are implemented at the same level (say, a cluster of postcodes, etc.).

To keep our notation simple, we consider the case of linking and estimating at the population level. In other words $S$ from the section 4 is such that $S = \Omega$, and therefore $|S| = |\Omega|$.

Denote the joint m-probability for a comparison pattern $\boldsymbol{\gamma} \in \Gamma$ as $m_{\boldsymbol{\gamma}}$. Using the conditional independence assumption (4) this probability can be computed as

\begin{equation}
m_{\boldsymbol{\gamma}} = \mathbb{P}(\boldsymbol{\gamma} \mid  M) = \prod_{v = 1} ^ {k} m_{v} ^ {\gamma_{v}}(1 - m_{v}) ^ {1 - {\gamma_{v}}}.
\end{equation}

Note that $p\cdot m_{\boldsymbol{\gamma}} = \mathbb{P}(\boldsymbol{\gamma} \cap M)$ (where $p$ is the proportion of matches among all pairs as defined in section 4) is the probability of observing a comparison pattern $\boldsymbol{\gamma}$ among the matches. Consider the following estimator of the number of matched cases for a comparison pattern $\boldsymbol{\gamma}$

\begin{equation}
\hat{n}_{11, \boldsymbol{\gamma}} = \hat{p} \cdot \hat{m}_{\boldsymbol{\gamma}} \cdot |\Omega|.
\end{equation}

Then the total number of matched cases can be estimated by

\begin{equation}
\hat{n}_{11} = \sum_{\boldsymbol{\gamma} \in \Gamma} \hat{p} \cdot \hat{m}_{\boldsymbol{\gamma}} \cdot |\Omega| = \hat{p} \cdot |\Omega|  \sum_{\boldsymbol{\gamma} \in \Gamma} \hat{m}_{\boldsymbol{\gamma}} = \hat{p} \cdot |\Omega|,
\end{equation}
since
\[\sum_{\boldsymbol{\gamma} \in \Gamma} \hat{m}_{\boldsymbol{\gamma}} = 1.\]

If we replace the number of matched cases, $n_{11}$, in (1) by its estimate (8) and notice that due to linking and estimating at the same level $n_{1+} = |L_{1}|$, $n_{+1} = |L_{2}|$, the population size estimator becomes

\begin{equation}
\hat{N}_{L} = \frac{n_{1+}n_{+1}}{\hat{n}_{11}} = \frac{n_{1+}n_{+1}}{\hat{p} \cdot |\Omega|} = \frac{|L_{1}| \cdot |L_{2}|}{\hat{p} \cdot |\Omega|} = \frac{|\Omega|}{\hat{p} \cdot |\Omega|} = \frac{1}{\hat{p}. }
\end{equation}

Note also that if $\hat{p}$ is the maximum likelihood estimator of $p$ then by the invariance property \cite{zehna66, pawitan13} $1 / \hat{p}$ is the maximum likelihood estimator of $1 / p$.

Hence, given that the population size and the linkage parameters are estimated at the same estimation stratum / block level, the function $f(p) = 1 / p$ of the linkage parameter $p$ is the estimator for $N$ and we call this estimator the LFDSE. When multiple estimation strata / linkage blocks are present, the above estimator is applied at each individual strata / block and resulting estimates can be summed up to obtain the population total.

Simplistically, $\hat{N}_{L}$ is easy to derive, since by definition $p = |M| / (|\Omega|) = |M| / (|L_{1}| \cdot |L_{2}|)$ and $1 / p = (|L_{1}| \cdot |L_{2}|) / |M|$ is the classical DSE. Thus, ensuring that the estimation and linkage take place at the same level is sufficient to allow this estimator to be used.

Since the analytical properties of the estimator $\hat{N}_{L}$ are not known at the moment, we use a small simulation study in the following section to explore the design based properties.

\section{Simulation study}
\label{simulation}
In this section we present and discuss the results of a small simulation study to assess the performance of the LFDSE, $\hat{N}_{L}$, and compare it to the simple DSE with perfect linkage, $\hat{N}$. Obviously, there is no need to simulate for the simple DSE since there are analytical expressions for its bias and variance \cite{wolter86}. Nevertheless, having simulated results for $\hat{N}$ makes the comparison easier and also gives an indication of the amount of random fluctuation for each scenario.

All the assumptions of the DSE and the linkage model hold in this study. We randomly generate two incomplete lists from the elements of the same population using the probabilities from scenarios below. We also generate the corresponding comparison outcomes for the resulting record pairs. We investigate the performance of $\hat{N}_{L}$ for the combinations of varying population size $N = \{150, 1000\}$, varying coverage probabilities $p_{l} = \{0.5, 0.7, 0.9\}$ of the lists $l = \{L_{1}, L_{2}\}$ and varying number of linkage variables $k = \{4, 6\}$.

In the case $k = 6$ we consider the following combinations of parameters:  $N = \{150, 1000\}$, all unique unordered  combinations of the coverage probabilities, the vector of  m-probabilities $\boldsymbol{m}_{6, 1} = (0.7, 0.75, 0.8, 0.85, 0.9, 0.95)$, the vector of u-probabilities $\boldsymbol{u}_{6, 1} = (0.001, 0.01, 0.05, 0.1, 0.15, 0.2)$ and its reversed variant $\boldsymbol{u}_{6, 1, r} = (0.2, 0.15, 0.1, 0.05, 0.01, 0.001)$. Here components $m_{v}$ and $u_{v}$ are such that $m_{v} = \mathbb{P}(\gamma_v = 1\mid M)$, $u_{v} = \mathbb{P}(\gamma_v = 1\mid U)$, $v = 1,\dots, k$. We also consider the vector of u-probabilities $\boldsymbol{u}_{6, 2} = (0.05, 0.1, 0.15, 0.2, 0.2, 0.25)$ and its reverse $\boldsymbol{u}_{6, 2, r} = (0.25, 0.2, 0.2, 0.15, 0.1, 0.05)$ in combination with $\boldsymbol{m}_{6, 1}$ and the coverage probability of 0.7 for the both sources. In the case $k = 4$ we consider the following parameters: $N = \{150, 1000\}$, combinations with equal coverage probabilities (to keep the number of scenarios low), the vector of m-probabilities $\boldsymbol{m}_{4, 1} = (0.7, 0.8, 0.9, 0.95)$, vectors of u-probabilities $\boldsymbol{u}_{4, 1} = (0.001, 0.01, 0.1, 0.2)$, $\boldsymbol{u}_{4, 2} = (0.01, 0.05, 0.1, 0.2)$ alongside their reversed versions $\boldsymbol{u}_{4, 1, r} = (0.2, 0.1, 0.01, 0.001)$ and $\boldsymbol{u}_{4, 2, r} = (0.2, 0.1, 0.05, 0.01)$. In addition, we consider the vector $\boldsymbol{u}_{4, 3} = (0.005, 0.01, 0.01, 0.03)$ and its reverse $\boldsymbol{u}_{4, 3, r} = (0.03, 0.01, 0.01, 0.005)$ in combination with $\boldsymbol{m}_{4, 1}$ and the coverage probability of 0.7 for both sources. Four more scenarios are included: $\boldsymbol{m}_{6, 2} = (0.9, 0.9, 0.9, 0.9, 0.9, 0.9)$, $\boldsymbol{u}_{6, 4} = (0.005, 0.005, 0.005, 0.005, 0.005, 0.005)$ when $k = 6$, and $\boldsymbol{m}_{4, 2} = (0.9, 0.9, 0.9, 0.9)$, $\boldsymbol{u}_{4, 4} = (0.005, 0.005, 0.005, 0.005)$ when $k = 4$ with $N = \{150, 1000\}$ and the coverage probability of 0.7 for the both sources. The total number of simulation scenarios is thus 60, the number of simulation iterations for each scenario is 100,000.

The quality measures include the relative bias (\textsc{rb}), the relative standard error (\textsc{rse}) and the relative root mean square error (\textsc{rrmse}):
\begin{align*}
	&\text{\textsc{rb}}(\hat{N}_{L}) = (\bar{\hat{N}}_{L} - N)/N,\\
	&\text{\textsc{rse}}(\hat{N}_{L}) = \sqrt{\text{var}(\hat{N}_{L})} / N,\\
	&\text{\textsc{rrmse}}(\hat{N}_{L}) = \sqrt{\text{var}(\hat{N}_{L}) + (\bar{\hat{N}}_{L} - N) ^ 2}/N
\end{align*}

We also report two additional measures facilitating the comparison between the DSE and the LFDSE. The first one is the ratio of the standard errors of $\hat{N}_L$ and $\hat{N}$. The second one is an approximate net linkage error (in this case, the absolute value of the difference between the number of false positive and the number of false negative errors, $\epsilon \doteq |\#(\text{FP}) - \#(\text{FN})|$) that would cause the \textsc{rrmse}($\hat{N}$) to be equal to the \textsc{rrmse}($\hat{N}_{L}$) obtained in the simulations. In other words, we are reporting $\epsilon$ such that
\[\text{\textsc{rrmse}}(\hat{N}_{\epsilon}) = \text{\textsc{rrmse}}(\hat{N}_{L})
\]
where 
\begin{equation}
\hat{N}_{\epsilon} = \frac{n_{1+}n_{+1}}{n_{11} \pm \epsilon}
\end{equation}
is the DSE when the linkage error is present. Assuming for simplicity that $\text{var}(\hat{N}) \approx \text{var}(\hat{N_{\epsilon}})$ we can calculate
\[\epsilon \approx \left|\frac{E(n_{1+}n_{+1})}{N + [\{\textsc{rrmse}(\hat{N}_{L})N\} ^ {2} - \text{var}(\hat{N})] ^ {1/2}} - E(n_{11}) \right|
\]

We report both the net number of erroneous cases, $\epsilon$, and the net error expressed as the percentage of the expected number of matches, $\epsilon / (Np_{1}p_{2}) \cdot 100\%$. A subset of the simulation results is presented in Table 1, results for all scenarios can be found in Appendix.

For a fixed pattern of $p_{1}, p_{2}, \boldsymbol{m}, \boldsymbol{u}$, as one would anticipate, the larger the population size, the smaller the relative bias and the relative standard error of $\hat{N}_{L}$. This is, of course, true for $\hat{N}$, but the DSE with perfect linkage outperforms $\hat{N}_{L}$ both in terms of bias and variance.

For a fixed pattern of $N, \boldsymbol{m}, \boldsymbol{u}$ varying coverage probabilities of the lists have similar effects on $\hat{N}$ and $\hat{N}_{L}$: the higher the coverage, the lower the relative bias and the relative error. Again, $\hat{N}$ outperforms $\hat{N}_{L}$ in both metrics.

\begin{table}
	\caption{Simulation results (subset)}
	\begin{center}
		{\scriptsize	
			\begin{tabular}{ r r r r r r r r r r r r r r r }
				\hline\hline
				&  &  &  &  &  & \multicolumn{2}{c}{\textsc{rb}\%} & \multicolumn{2}{c}{\textsc{rse}\%} & \multicolumn{2}{c}{\textsc{rrmse}\%} &  &  \multicolumn{2}{c}{net error} \\ 			
				$\#$ & $N$ & $p_{1}$ & $p_{2}$ & $\boldsymbol{m}$ & $\boldsymbol{u}$ & $\hat{N}$ & $\hat{N}_{L}$ & $\hat{N}$ & $\hat{N}_{L}$ & $\hat{N}$ & $\hat{N}_{L}$ & ratio & $\epsilon$ & \% \\ \hline
				1 	&	1000	&	0.5	&	0.5	&	 $\boldsymbol{m}_{6, 1}$ 	&	 $\boldsymbol{u}_{6, 1}$ 	&	0.09	&	0.15	&	3.20	&	4.71	&	3.20	&	4.72	&	1.47	&	8.37	&	3.35	\\
				2 	&	1000	&	0.5	&	0.5	&	 $\boldsymbol{m}_{6, 1}$ 	&	 $\boldsymbol{u}_{6, 1, r}$ 	&	0.11	&	0.13	&	3.19	&	3.82	&	3.19	&	3.82	&	1.20	&	5.15	&	2.06	\\
				3 	&	1000	&	0.5	&	0.7	&	 $\boldsymbol{m}_{6, 1}$ 	&	 $\boldsymbol{u}_{6, 1}$ 	&	0.04	&	0.09	&	2.09	&	3.59	&	2.09	&	3.59	&	1.72	&	9.94	&	2.84	\\
				4 	&	1000	&	0.5	&	0.7	&	 $\boldsymbol{m}_{6, 1}$ 	&	 $\boldsymbol{u}_{6, 1, r}$ 	&	0.05	&	0.07	&	2.08	&	2.74	&	2.09	&	2.74	&	1.31	&	6.10	&	1.74	\\
				5 	&	1000	&	0.5	&	0.9	&	 $\boldsymbol{m}_{6, 1}$ 	&	 $\boldsymbol{u}_{6, 1}$ 	&	0.02	&	0.04	&	1.06	&	2.77	&	1.06	&	2.77	&	2.62	&	11.22	&	2.49	\\
				6 	&	1000	&	0.5	&	0.9	&	 $\boldsymbol{m}_{6, 1}$ 	&	 $\boldsymbol{u}_{6, 1, r}$ 	&	0.01	&	0.03	&	1.06	&	1.88	&	1.06	&	1.88	&	1.77	&	6.86	&	1.52	\\
				7 	&	1000	&	0.7	&	0.7	&	 $\boldsymbol{m}_{6, 1}$ 	&	 $\boldsymbol{u}_{6, 1}$ 	&	0.01	&	0.04	&	1.36	&	2.80	&	1.36	&	2.80	&	2.06	&	11.73	&	2.39	\\
				8 	&	1000	&	0.7	&	0.7	&	 $\boldsymbol{m}_{6, 1}$ 	&	 $\boldsymbol{u}_{6, 1, r}$ 	&	0.02	&	0.02	&	1.36	&	2.01	&	1.36	&	2.01	&	1.48	&	7.17	&	1.46	\\
				9 	&	1000	&	0.7	&	0.9	&	 $\boldsymbol{m}_{6, 1}$ 	&	 $\boldsymbol{u}_{6, 1}$ 	&	0.00	&	0.02	&	0.69	&	2.27	&	0.69	&	2.27	&	3.28	&	13.35	&	2.12	\\
				10 	&	1000	&	0.7	&	0.9	&	 $\boldsymbol{m}_{6, 1}$ 	&	 $\boldsymbol{u}_{6, 1, r}$ 	&	0.01	&	0.02	&	0.69	&	1.48	&	0.69	&	1.48	&	2.16	&	8.18	&	1.30	\\
				11 	&	1000	&	0.9	&	0.9	&	 $\boldsymbol{m}_{6, 1}$ 	&	 $\boldsymbol{u}_{6, 1}$ 	&	0.00	&	0.02	&	0.35	&	1.94	&	0.35	&	1.94	&	5.53	&	15.17	&	1.87	\\
				12 	&	1000	&	0.9	&	0.9	&	 $\boldsymbol{m}_{6, 1}$ 	&	 $\boldsymbol{u}_{6, 1, r}$ 	&	0.00	&	0.01	&	0.35	&	1.21	&	0.35	&	1.21	&	3.43	&	9.25	&	1.14	\\
				13 	&	150	&	0.5	&	0.5	&	 $\boldsymbol{m}_{6, 1}$ 	&	 $\boldsymbol{u}_{6, 1}$ 	&	0.72	&	0.93	&	8.53	&	10.69	&	8.56	&	10.73	&	1.25	&	2.29	&	6.11	\\
				14 	&	150	&	0.5	&	0.5	&	 $\boldsymbol{m}_{6, 1}$ 	&	 $\boldsymbol{u}_{6, 1, r}$ 	&	0.69	&	0.71	&	8.56	&	9.38	&	8.59	&	9.40	&	1.10	&	1.40	&	3.74	\\
				15 	&	150	&	0.5	&	0.7	&	 $\boldsymbol{m}_{6, 1}$ 	&	 $\boldsymbol{u}_{6, 1}$ 	&	0.29	&	0.45	&	5.53	&	7.61	&	5.54	&	7.62	&	1.38	&	2.62	&	4.99	\\
				16 	&	150	&	0.5	&	0.7	&	 $\boldsymbol{m}_{6, 1}$ 	&	 $\boldsymbol{u}_{6, 1, r}$ 	&	0.30	&	0.33	&	5.51	&	6.36	&	5.52	&	6.37	&	1.15	&	1.62	&	3.09	\\
				17 	&	150	&	0.5	&	0.9	&	 $\boldsymbol{m}_{6, 1}$ 	&	 $\boldsymbol{u}_{6, 1}$ 	&	0.06	&	0.15	&	2.77	&	5.26	&	2.77	&	5.27	&	1.90	&	2.89	&	4.29	\\
				18 	&	150	&	0.5	&	0.9	&	 $\boldsymbol{m}_{6, 1}$ 	&	 $\boldsymbol{u}_{6, 1, r}$ 	&	0.08	&	0.11	&	2.78	&	3.89	&	2.78	&	3.89	&	1.40	&	1.79	&	2.65	\\
				19 	&	150	&	0.7	&	0.7	&	 $\boldsymbol{m}_{6, 1}$ 	&	 $\boldsymbol{u}_{6, 1}$ 	&	0.12	&	0.22	&	3.56	&	5.60	&	3.57	&	5.60	&	1.57	&	3.05	&	4.15	\\
				20 	&	150	&	0.7	&	0.7	&	 $\boldsymbol{m}_{6, 1}$ 	&	 $\boldsymbol{u}_{6, 1, r}$ 	&	0.13	&	0.17	&	3.55	&	4.40	&	3.55	&	4.40	&	1.24	&	1.86	&	2.54	\\
				21 	&	150	&	0.7	&	0.9	&	 $\boldsymbol{m}_{6, 1}$ 	&	 $\boldsymbol{u}_{6, 1}$ 	&	0.04	&	0.09	&	1.80	&	4.16	&	1.80	&	4.17	&	2.32	&	3.42	&	3.62	\\
				22 	&	150	&	0.7	&	0.9	&	 $\boldsymbol{m}_{6, 1}$ 	&	 $\boldsymbol{u}_{6, 1, r}$ 	&	0.03	&	0.08	&	1.80	&	2.91	&	1.80	&	2.91	&	1.62	&	2.11	&	2.24	\\
				23 	&	150	&	0.9	&	0.9	&	 $\boldsymbol{m}_{6, 1}$ 	&	 $\boldsymbol{u}_{6, 1}$ 	&	0.01	&	0.06	&	0.91	&	3.40	&	0.91	&	3.40	&	3.73	&	3.86	&	3.17	\\
				24 	&	150	&	0.9	&	0.9	&	 $\boldsymbol{m}_{6, 1}$ 	&	 $\boldsymbol{u}_{6, 1, r}$ 	&	0.01	&	0.04	&	0.91	&	2.18	&	0.91	&	2.19	&	2.39	&	2.37	&	1.95	\\
				25 	&	1000	&	0.5	&	0.5	&	 $\boldsymbol{m}_{4, 1}$ 	&	 $\boldsymbol{u}_{4, 1}$ 	&	0.11	&	0.31	&	3.18	&	6.63	&	3.18	&	6.63	&	2.09	&	13.76	&	5.50	\\
				27 	&	1000	&	0.5	&	0.5	&	 $\boldsymbol{m}_{4, 1}$ 	&	 $\boldsymbol{u}_{4, 2}$ 	&	0.09	&	1.31	&	3.18	&	12.94	&	3.18	&	13.01	&	4.07	&	28.00	&	11.20	\\
				29 	&	1000	&	0.7	&	0.7	&	 $\boldsymbol{m}_{4, 1}$ 	&	 $\boldsymbol{u}_{4, 1}$ 	&	0.02	&	0.14	&	1.36	&	4.31	&	1.36	&	4.31	&	3.18	&	19.27	&	3.93	\\
				30 	&	1000	&	0.7	&	0.7	&	 $\boldsymbol{m}_{4, 1}$ 	&	 $\boldsymbol{u}_{4, 1, r}$ 	&	0.01	&	-0.12	&	1.36	&	4.32	&	1.36	&	4.32	&	3.18	&	19.32	&	3.94	\\
				31 	&	1000	&	0.7	&	0.7	&	 $\boldsymbol{m}_{4, 1}$ 	&	 $\boldsymbol{u}_{4, 2}$ 	&	0.02	&	0.68	&	1.36	&	8.91	&	1.36	&	8.93	&	6.55	&	39.75	&	8.11	\\
				39 	&	150	&	0.5	&	0.5	&	 $\boldsymbol{m}_{4, 1}$ 	&	 $\boldsymbol{u}_{4, 2}$ 	&	0.72	&	1.23	&	8.56	&	17.91	&	8.59	&	17.95	&	2.09	&	5.11	&	13.63	\\
				49 	&	1000	&	0.7	&	0.7	&	 $\boldsymbol{m}_{6, 1}$ 	&	 $\boldsymbol{u}_{6, 2}$ 	&	0.02	&	0.41	&	1.36	&	8.47	&	1.36	&	8.48	&	6.22	&	37.83	&	7.72	\\
				53 	&	1000	&	0.7	&	0.7	&	 $\boldsymbol{m}_{4, 1}$ 	&	 $\boldsymbol{u}_{4, 3}$ 	&	0.02	&	0.04	&	1.36	&	2.72	&	1.36	&	2.72	&	2.00	&	11.30	&	2.31	\\
				57 	&	1000	&	0.7	&	0.7	&	 $\boldsymbol{m}_{6, 2}$ 	&	 $\boldsymbol{u}_{6, 4}$ 	&	0.02	&	0.02	&	1.36	&	1.38	&	1.36	&	1.38	&	1.02	&	1.31	&	0.27	\\
				58 	&	150	&	0.7	&	0.7	&	 $\boldsymbol{m}_{6, 2}$ 	&	 $\boldsymbol{u}_{6, 4}$ 	&	0.14	&	0.14	&	3.56	&	3.59	&	3.56	&	3.60	&	1.01	&	0.38	&	0.52	\\
				59 	&	1000	&	0.7	&	0.7	&	 $\boldsymbol{m}_{4, 2}$ 	&	 $\boldsymbol{u}_{4, 4}$ 	&	0.02	&	0.01	&	1.36	&	1.79	&	1.36	&	1.79	&	1.32	&	5.64	&	1.15	\\
				
				\hline
			\end{tabular}
		}
	\end{center}
\end{table}

Arguably, the vectors of m- and u-probabilities have the most interesting effect on $\hat{N}_{L}$. In order to keep the number of simulation scenarios computationally manageable and interpretable, we did not consider a wide range of m- and u-probability vectors. In addition, unlike other simulation parameters, there is no straightforward way to order the probability vectors, which makes the comparison difficult. With this limited simulation some uncertainty remains around the exact relationship between the m- and u-probabilities and the quality of estimates produced by $\hat{N}_{L}$. Intuitively, the higher the m-probabilities and the lower the u-probabilities (that is, the higher is discriminatory power of the linkage variables), the more accurate the estimator will be. It is also expected that the more linkage variables with high or moderate discriminatory power are available, the higher the accuracy will be. For fixed patterns of $N, p_{1}, p_{2}$ and $\boldsymbol{m}$, we can see in Table 1 that, indeed, the smaller the u-probabilities, the lower the relative bias and the relative standard error. For example, compare scenario 7 with 49, 29 with 53 and 29 with 31. Increasing the m-probabilities and simultaneously decreasing the u-probabilities improves the accuracy, for example compare 7 with 57, and 29 with 59. Having more linkage variables may lead to increased precision when u-probabilities are approximately 'the same', compare 7 with 29. Having fewer linkage variables, but with 'higher discriminatory power' (lower u-probabilities / higher m-probabilities) may result in higher accuracy, compare 7 with 53 or 59. As the u-probabilities increase, the quality of the estimates decreases both in terms of the relative bias and the relative error: compare 25 with 27. The accuracy varies with ways of pairing the components of the probability vectors: 7 with 8, 29 with 30. Interestingly, for the cases with 4 linkage variables and moderately high u-probabilities, the relative bias may switch from positive to negative depending on the pairing of the m- and u-probabilities: compare 27 with 28.

Comparing $\hat{N}_{L}$ with $\hat{N}$, we see that the proposed estimator is nearly equivalent to the standard DSE if the m-probabilities are all high, the u-probabilities are all low and there are sufficiently many linkage variables (57 and 58). Otherwise, the ratio of standard errors is between 1.5 and 3 for the majority of scenarios considered, but could be as high as 20.

It might be more informative to compare $\hat{N}_{L}$ with $\hat{N}$ on the basis of the net linkage error that would induce sufficient bias in the estimation to make the \textsc{rrmse} of $\hat{N}$ equal to the \textsc{rrmse} of $\hat{N}_{L}$ (two final columns in Table 1). We can see again, that if the number of linkage variables and their quality is very high, $\hat{N}_{L}$ and $\hat{N}$ are virtually equivalent with the expected net errors of 1.31 and 0.39 for the population sizes 1000 and 150, respectively. When the linkage variables are not all of such exceptional quality, the net linkage error needed to reduce the performance of $\hat{N}$ so that it is similar to the performance of $\hat{N}_{L}$ can vary substantially. Among the scenarios considered the net percentage error can be as small as 1.14\% (net expected error of 9.25 out of expected 810 matches) when $N = 1000$, $k = 6$ (scenario 12), or 1.95\% (net expected error of 2.37 out of expected 121.5 matches) when $N = 150$, $k = 6$ (scenario 24). On the other hand, this error can be as large as 11.20\% (net expected error of 30.00 out of expected 250 matches) when $N = 1000$, $k = 4$ (scenario 27), or 13.63\% (net expected error of 5.11 out of expected 37.5 matches) when $N = 150$, $k = 4$ (scenario 39).

In practice, the choice between $\hat{N}_{L}$ and $\hat{N}$ would largely depend on the number of linkage variables, their quality (high m-probabilities, low u-probabilities) and the level of admissible linkage errors for $\hat{N}$.

\section{Variance estimation}
\label{variance}
In this section we propose a parametric bootstrap approach to estimating the variance of $\hat{N}_{L}$ and we also assess its performance. The algorithm is as follows:

\begin{enumerate} 
\item For the observed comparison patterns of two data sets $L_{1}$ and $L_{2}$, use the method described in section 4 to estimate $\hat{p}, \boldsymbol{\hat{m}}, \boldsymbol{\hat{u}}$. Then compute $\hat{N}_{L}$;
\item Use $\hat{N}_{L}$ and the observed marginal counts $n_{1+}, n_{+1}$ of $L_{1}$ and $L_{2}$ to estimate the marginal probabilities: $\hat{p}_{1+} = n_{1+} / \hat{N}_L, \hat{p}_{+1} = n_{+1} / \hat{N}_L$;
\item Estimate the four cell probabilities under the independence assumption: $\hat{p}_{11} = \hat{p}_{1+}\hat{p}_{+1}$, etc.;
\item Use the cell probabilities estimated in step 3 to draw counts from the multinomial distribution $M(\hat{N}_L, \hat{p}_{11}, \hat{p}_{10}, \hat{p}_{01}, \hat{p}_{00})$;
\item Calculate the marginal counts $n_{1+}^{*}, n_{+1}^{*}$ and the match count $n_{11}^{*}$ from the cells generated in step 4;
\item Use $\boldsymbol{\hat{m}}, \boldsymbol{\hat{u}}$ estimated in step 1 to compute the joint m- and u-probabilities for each comparison pattern $\boldsymbol{\gamma}$ as in (6);
\item Draw the comparison outcomes for the M-set from the multinomial distribution with parameters: $n_{11}^{*}$ and the vector of joint m-probabilities obtained in step 6;
\item Draw the comparison outcomes for the U-set from the multinomial distribution with parameters: $n_{1+}^{*}n_{+1}^{*} - n_{11}^{*}$ and the vector of joint u-probabilities obtained in step 6;
\item Combine the patterns from steps 7 and 8 into a single data set;
\item Use the data set from step 9 to estimate $\hat{p}^{*}$ and compute ${\hat{N}_{L}}^{*} = 1 / \hat{p}^{*}$;
\item Iterate steps 2 to 10 a sufficient number of times.
\end{enumerate}

We test the performance of the estimator on a subset of the scenarios considered in section 6. We generate 1000 pairs of $L_{1}$ and $L_{2}$ and for each pair we perform 1000 iterations of steps 2 to 10. For a number of cases, we generate 3000 pairs of $L_{1}$,  $L_{2}$ and then perform 1000 iterations of steps 2 to 10 (not presented). This is to confirm that the results are not affected substantially by random fluctuations. The simulated (non-bootstrap) mean relative standard error of $\hat{N}_{L}$, the mean relative standard error of $\hat{N}_{L} ^ {*}$ estimated by the above bootstrap method and the empirical (bootstrap) coverage of the 95\% confidence interval, $95\%CI_{cov}$, are reported in Table 2.

As can be seen, the majority of the relative standard errors estimated by the above bootstrap procedure are close to those obtained in the simulation study described in section 6 and the 95\% bootstrap confidence intervals have coverage probabilities close to the nominal value. However, in some cases the variance estimation procedure does not perform very well. This occurs with four linkage variables when all the u-probabilities are high. In these scenarios the variance is overestimated and the coverage probability exceeds the nominal level. These results were confirmed by increasing the number of iterations at step 1 to 3000. Further work is needed to understand this behaviour.

\begin{table}
	\caption{Variance estimation results (all statistics in percentages)}
	
	\begin{center}	
		{\scriptsize	
			\begin{tabular}{ r r r r r r r r }
				\hline\hline
				$N$ & $p_{1}$ & $p_{2}$ & $\boldsymbol{m}$ & $\boldsymbol{u}$ & $\textsc{rse}(\hat{N}_{L})$ & $\textsc{rse}(\hat{N}_{L} ^ {*})$ & $95\%CI_{cov}$ \\ 
				\hline
				
				1000 & 0.5 & 0.5 & $\boldsymbol{m}_{6, 1}$ & $\boldsymbol{u}_{6, 1}$ & 4.61 & 4.71 & 96.1 \\ 
				1000 & 0.7 & 0.9 & $\boldsymbol{m}_{6, 1}$ & $\boldsymbol{u}_{6, 1}$ & 2.28 & 2.26 & 94.8 \\ 
				1000 & 0.9 & 0.9 & $\boldsymbol{m}_{6, 1}$ & $\boldsymbol{u}_{6, 1}$ & 1.95 &  1.94 & 95.6 \\ 
				150 & 0.5 & 0.5 & $\boldsymbol{m}_{6, 1}$ & $\boldsymbol{u}_{6, 1}$ & 10.22 & 10.93 & 96.4 \\ 
				150 & 0.7 & 0.9 & $\boldsymbol{m}_{6, 1}$ & $\boldsymbol{u}_{6, 1}$ & 4.31 &  4.18 & 94.9 \\ 
				150 & 0.9 & 0.9 & $\boldsymbol{m}_{6, 1}$ & $\boldsymbol{u}_{6, 1}$ & 3.44 &  3.38 & 95.2 \\ 
				1000 & 0.5 & 0.5 & $\boldsymbol{m}_{4, 1}$ & $\boldsymbol{u}_{4, 1}$ & 6.74 &  6.66 & 94.2 \\ 
				1000 & 0.7 & 0.9 & $\boldsymbol{m}_{4, 1}$ & $\boldsymbol{u}_{4, 2}$ & 7.45 &  7.86 & 98.6 \\ 
				1000 & 0.9 & 0.9 & $\boldsymbol{m}_{4, 1}$ & $\boldsymbol{u}_{4, 1}$ & 3.19 &  3.19 & 94.9 \\ 
				150 & 0.5 & 0.5 & $\boldsymbol{m}_{4, 1}$ & $\boldsymbol{u}_{4, 1}$ & 13.69 & 13.94 & 95.8 \\ 
				150 & 0.7 & 0.9 & $\boldsymbol{m}_{4, 1}$ & $\boldsymbol{u}_{4, 2}$ & 8.89 &  9.82 & 98.4 \\ 
				150 & 0.9 & 0.9 & $\boldsymbol{m}_{4, 1}$ & $\boldsymbol{u}_{4, 1}$ & 5.22 &  5.41 & 96.5 \\ 
				\hline
				
			\end{tabular}
		}
	\end{center}
\end{table}

\section{Discussion and further work}
\label{discussion}
The DSE relies upon high quality linkage. Record linkage for population size estimation is often a complex (because of the need to meet the 1-1 matching constraint) and resource intensive (because of the need to resolve potential links clerically) process. Usually, the practical difficulty of achieving perfect linkage is recognised by setting some admissible level for linkage errors and accepting the consequent small bias. However, procedures to adjust the DSE to account for the linkage have been developed \cite{dingfienb94, dewolfetal18}, although they require independent estimates of the linkage errors. 

In this paper, a linkage free DSE, $\hat{N}_{L}$, and a corresponding parametric bootstrap variance estimation procedure are proposed. Unlike the basic DSE, the LFDSE does not require the record pairs to be classified as links and non-links. It requires only parameter estimation for the Fellegi--Sunter probabilistic linkage model and thus simplifies the estimation process by eliminating the 1-1 match constraint and clerical review of unlinked record pairs. But of course, this method additionally requires the linkage model assumptions to hold.

The basic DSE with perfectly linked data is more accurate than the LFDSE, but as the number and discriminatory power of the linkage variables increase, the performance of the LFDSE approaches the performance of the DSE. When the linkage is not perfect, the performance of the LFDSE depends on many factors. When the linkage error exceeds certain levels (which may be quite low) the LFDSE outperforms the DSE in \textsc{rrmse} as well as being much easier and cheaper to implement.

The LFDSE requires that the estimation and linkage model are both specified at the same stratum / block level. There are some known applications of the DSE where the linkage blocks and estimation strata are closely related: the coverage estimation for the 2011 Census of England and Wales \cite{brownetal18} and a combinations of the LFDSE and the ratio estimator is one situation where this is an interesting model in practice.

There are several areas where further research on the LFDSE is needed. One strand is relaxing or dealing with departures from the remaining assumptions. For instance, relaxing the conditional independence assumption \cite{larsenrubin01}. Another strand of research is to investigate the performance of the LFDSE and adapt it for situations when values for a certain linkage variable are not uniformly distributed (say, some surnames are more common than other). Yet one more strand is to test the performance of the proposed estimator on real examples. And of course, understanding how the m- and u-probabilities are related to the performance of the LFDSE is of both theoretical interest and practical use.

\bibliographystyle{plain}
\bibliography{lfdse_v1_4b}

\appendix

\section*{Appendix}
\subsection*{Complete table of simulation results}
	
{\scriptsize	
	\begin{longtable}{ r r r r r r r r r r r r r r r }
		\hline\hline
		&  &  &  &  &  & \multicolumn{2}{c}{\textsc{rb}\%} & \multicolumn{2}{c}{\textsc{rse}\%} & \multicolumn{2}{c}{\textsc{rrmse}\%} &  &  \multicolumn{2}{c}{net error} \\ 			
		$\#$ & $N$ & $p_{1}$ & $p_{2}$ & $\boldsymbol{m}$ & $\boldsymbol{u}$ & $\hat{N}$ & $\hat{N}_{L}$ & $\hat{N}$ & $\hat{N}_{L}$ & $\hat{N}$ & $\hat{N}_{L}$ & ratio & $\epsilon$ & \% \\ \hline
		1 	&	1000	&	0.5	&	0.5	&	 $\boldsymbol{m}_{6, 1}$ 	&	 $\boldsymbol{u}_{6, 1}$ 	&	0.09	&	0.15	&	3.20	&	4.71	&	3.20	&	4.72	&	1.47	&	8.37	&	3.35	\\
		2 	&	1000	&	0.5	&	0.5	&	 $\boldsymbol{m}_{6, 1}$ 	&	 $\boldsymbol{u}_{6, 1, r}$ 	&	0.11	&	0.13	&	3.19	&	3.82	&	3.19	&	3.82	&	1.20	&	5.15	&	2.06	\\
		3 	&	1000	&	0.5	&	0.7	&	 $\boldsymbol{m}_{6, 1}$ 	&	 $\boldsymbol{u}_{6, 1}$ 	&	0.04	&	0.09	&	2.09	&	3.59	&	2.09	&	3.59	&	1.72	&	9.94	&	2.84	\\
		4 	&	1000	&	0.5	&	0.7	&	 $\boldsymbol{m}_{6, 1}$ 	&	 $\boldsymbol{u}_{6, 1, r}$ 	&	0.05	&	0.07	&	2.08	&	2.74	&	2.09	&	2.74	&	1.31	&	6.10	&	1.74	\\
		5 	&	1000	&	0.5	&	0.9	&	 $\boldsymbol{m}_{6, 1}$ 	&	 $\boldsymbol{u}_{6, 1}$ 	&	0.02	&	0.04	&	1.06	&	2.77	&	1.06	&	2.77	&	2.62	&	11.22	&	2.49	\\
		6 	&	1000	&	0.5	&	0.9	&	 $\boldsymbol{m}_{6, 1}$ 	&	 $\boldsymbol{u}_{6, 1, r}$ 	&	0.01	&	0.03	&	1.06	&	1.88	&	1.06	&	1.88	&	1.77	&	6.86	&	1.52	\\
		7 	&	1000	&	0.7	&	0.7	&	 $\boldsymbol{m}_{6, 1}$ 	&	 $\boldsymbol{u}_{6, 1}$ 	&	0.01	&	0.04	&	1.36	&	2.80	&	1.36	&	2.80	&	2.06	&	11.73	&	2.39	\\
		8 	&	1000	&	0.7	&	0.7	&	 $\boldsymbol{m}_{6, 1}$ 	&	 $\boldsymbol{u}_{6, 1, r}$ 	&	0.02	&	0.02	&	1.36	&	2.01	&	1.36	&	2.01	&	1.48	&	7.17	&	1.46	\\
		9 	&	1000	&	0.7	&	0.9	&	 $\boldsymbol{m}_{6, 1}$ 	&	 $\boldsymbol{u}_{6, 1}$ 	&	0.00	&	0.02	&	0.69	&	2.27	&	0.69	&	2.27	&	3.28	&	13.35	&	2.12	\\
		10 	&	1000	&	0.7	&	0.9	&	 $\boldsymbol{m}_{6, 1}$ 	&	 $\boldsymbol{u}_{6, 1, r}$ 	&	0.01	&	0.02	&	0.69	&	1.48	&	0.69	&	1.48	&	2.16	&	8.18	&	1.30	\\
		11 	&	1000	&	0.9	&	0.9	&	 $\boldsymbol{m}_{6, 1}$ 	&	 $\boldsymbol{u}_{6, 1}$ 	&	0.00	&	0.02	&	0.35	&	1.94	&	0.35	&	1.94	&	5.53	&	15.17	&	1.87	\\
		12 	&	1000	&	0.9	&	0.9	&	 $\boldsymbol{m}_{6, 1}$ 	&	 $\boldsymbol{u}_{6, 1, r}$ 	&	0.00	&	0.01	&	0.35	&	1.21	&	0.35	&	1.21	&	3.43	&	9.25	&	1.14	\\
		13 	&	150	&	0.5	&	0.5	&	 $\boldsymbol{m}_{6, 1}$ 	&	 $\boldsymbol{u}_{6, 1}$ 	&	0.72	&	0.93	&	8.53	&	10.69	&	8.56	&	10.73	&	1.25	&	2.29	&	6.11	\\
		14 	&	150	&	0.5	&	0.5	&	 $\boldsymbol{m}_{6, 1}$ 	&	 $\boldsymbol{u}_{6, 1, r}$ 	&	0.69	&	0.71	&	8.56	&	9.38	&	8.59	&	9.40	&	1.10	&	1.40	&	3.74	\\
		15 	&	150	&	0.5	&	0.7	&	 $\boldsymbol{m}_{6, 1}$ 	&	 $\boldsymbol{u}_{6, 1}$ 	&	0.29	&	0.45	&	5.53	&	7.61	&	5.54	&	7.62	&	1.38	&	2.62	&	4.99	\\
		16 	&	150	&	0.5	&	0.7	&	 $\boldsymbol{m}_{6, 1}$ 	&	 $\boldsymbol{u}_{6, 1, r}$ 	&	0.30	&	0.33	&	5.51	&	6.36	&	5.52	&	6.37	&	1.15	&	1.62	&	3.09	\\
		17 	&	150	&	0.5	&	0.9	&	 $\boldsymbol{m}_{6, 1}$ 	&	 $\boldsymbol{u}_{6, 1}$ 	&	0.06	&	0.15	&	2.77	&	5.26	&	2.77	&	5.27	&	1.90	&	2.89	&	4.29	\\
		18 	&	150	&	0.5	&	0.9	&	 $\boldsymbol{m}_{6, 1}$ 	&	 $\boldsymbol{u}_{6, 1, r}$ 	&	0.08	&	0.11	&	2.78	&	3.89	&	2.78	&	3.89	&	1.40	&	1.79	&	2.65	\\
		19 	&	150	&	0.7	&	0.7	&	 $\boldsymbol{m}_{6, 1}$ 	&	 $\boldsymbol{u}_{6, 1}$ 	&	0.12	&	0.22	&	3.56	&	5.60	&	3.57	&	5.60	&	1.57	&	3.05	&	4.15	\\
		20 	&	150	&	0.7	&	0.7	&	 $\boldsymbol{m}_{6, 1}$ 	&	 $\boldsymbol{u}_{6, 1, r}$ 	&	0.13	&	0.17	&	3.55	&	4.40	&	3.55	&	4.40	&	1.24	&	1.86	&	2.54	\\
		21 	&	150	&	0.7	&	0.9	&	 $\boldsymbol{m}_{6, 1}$ 	&	 $\boldsymbol{u}_{6, 1}$ 	&	0.04	&	0.09	&	1.80	&	4.16	&	1.80	&	4.17	&	2.32	&	3.42	&	3.62	\\
		22 	&	150	&	0.7	&	0.9	&	 $\boldsymbol{m}_{6, 1}$ 	&	 $\boldsymbol{u}_{6, 1, r}$ 	&	0.03	&	0.08	&	1.80	&	2.91	&	1.80	&	2.91	&	1.62	&	2.11	&	2.24	\\
		23 	&	150	&	0.9	&	0.9	&	 $\boldsymbol{m}_{6, 1}$ 	&	 $\boldsymbol{u}_{6, 1}$ 	&	0.01	&	0.06	&	0.91	&	3.40	&	0.91	&	3.40	&	3.73	&	3.86	&	3.17	\\
		24 	&	150	&	0.9	&	0.9	&	 $\boldsymbol{m}_{6, 1}$ 	&	 $\boldsymbol{u}_{6, 1, r}$ 	&	0.01	&	0.04	&	0.91	&	2.18	&	0.91	&	2.19	&	2.39	&	2.37	&	1.95	\\
		25 	&	1000	&	0.5	&	0.5	&	 $\boldsymbol{m}_{4, 1}$ 	&	 $\boldsymbol{u}_{4, 1}$ 	&	0.11	&	0.31	&	3.18	&	6.63	&	3.18	&	6.63	&	2.09	&	13.76	&	5.50	\\
		26 	&	1000	&	0.5	&	0.5	&	 $\boldsymbol{m}_{4, 1}$ 	&	 $\boldsymbol{u}_{4, 1, r}$ 	&	0.09	&	-0.34	&	3.17	&	6.21	&	3.18	&	6.22	&	1.96	&	12.70	&	5.08	\\
		27 	&	1000	&	0.5	&	0.5	&	 $\boldsymbol{m}_{4, 1}$ 	&	 $\boldsymbol{u}_{4, 2}$ 	&	0.09	&	1.31	&	3.18	&	12.94	&	3.18	&	13.01	&	4.07	&	28.00	&	11.20	\\
		28 	&	1000	&	0.5	&	0.5	&	 $\boldsymbol{m}_{4, 1}$ 	&	 $\boldsymbol{u}_{4, 2, r}$ 	&	0.08	&	-2.03	&	3.18	&	11.13	&	3.18	&	11.31	&	3.50	&	24.49	&	9.79	\\
		29 	&	1000	&	0.7	&	0.7	&	 $\boldsymbol{m}_{4, 1}$ 	&	 $\boldsymbol{u}_{4, 1}$ 	&	0.02	&	0.14	&	1.36	&	4.31	&	1.36	&	4.31	&	3.18	&	19.27	&	3.93	\\
		30 	&	1000	&	0.7	&	0.7	&	 $\boldsymbol{m}_{4, 1}$ 	&	 $\boldsymbol{u}_{4, 1, r}$ 	&	0.01	&	-0.12	&	1.36	&	4.32	&	1.36	&	4.32	&	3.18	&	19.32	&	3.94	\\
		31 	&	1000	&	0.7	&	0.7	&	 $\boldsymbol{m}_{4, 1}$ 	&	 $\boldsymbol{u}_{4, 2}$ 	&	0.02	&	0.68	&	1.36	&	8.91	&	1.36	&	8.93	&	6.55	&	39.75	&	8.11	\\
		32 	&	1000	&	0.7	&	0.7	&	 $\boldsymbol{m}_{4, 1}$ 	&	 $\boldsymbol{u}_{4, 2, r}$ 	&	0.02	&	-1.11	&	1.36	&	8.26	&	1.36	&	8.33	&	6.08	&	37.22	&	7.60	\\
		33 	&	1000	&	0.9	&	0.9	&	 $\boldsymbol{m}_{4, 1}$ 	&	 $\boldsymbol{u}_{4, 1}$ 	&	0.00	&	0.05	&	0.35	&	3.20	&	0.35	&	3.20	&	9.06	&	24.96	&	3.08	\\
		34 	&	1000	&	0.9	&	0.9	&	 $\boldsymbol{m}_{4, 1}$ 	&	 $\boldsymbol{u}_{4, 1, r}$ 	&	0.00	&	-0.02	&	0.35	&	3.36	&	0.35	&	3.36	&	9.53	&	26.17	&	3.23	\\
		35 	&	1000	&	0.9	&	0.9	&	 $\boldsymbol{m}_{4, 1}$ 	&	 $\boldsymbol{u}_{4, 2}$ 	&	0.00	&	0.39	&	0.35	&	6.81	&	0.35	&	6.82	&	19.39	&	51.62	&	6.37	\\
		36 	&	1000	&	0.9	&	0.9	&	 $\boldsymbol{m}_{4, 1}$ 	&	 $\boldsymbol{u}_{4, 2, r}$ 	&	0.00	&	-0.72	&	0.35	&	6.68	&	0.35	&	6.72	&	18.95	&	50.96	&	6.29	\\
		37 	&	150	&	0.5	&	0.5	&	 $\boldsymbol{m}_{4, 1}$ 	&	 $\boldsymbol{u}_{4, 1}$ 	&	0.69	&	1.07	&	8.53	&	13.52	&	8.56	&	13.56	&	1.58	&	3.58	&	9.53	\\
		38 	&	150	&	0.5	&	0.5	&	 $\boldsymbol{m}_{4, 1}$ 	&	 $\boldsymbol{u}_{4, 1, r}$ 	&	0.70	&	0.00	&	8.54	&	11.06	&	8.57	&	11.06	&	1.29	&	2.46	&	6.56	\\
		39 	&	150	&	0.5	&	0.5	&	 $\boldsymbol{m}_{4, 1}$ 	&	 $\boldsymbol{u}_{4, 2}$ 	&	0.72	&	1.23	&	8.56	&	17.91	&	8.59	&	17.95	&	2.09	&	5.11	&	13.63	\\
		40 	&	150	&	0.5	&	0.5	&	 $\boldsymbol{m}_{4, 1}$ 	&	 $\boldsymbol{u}_{4, 2, r}$ 	&	0.71	&	-1.69	&	8.58	&	15.22	&	8.61	&	15.31	&	1.77	&	4.22	&	11.25	\\
		41 	&	150	&	0.7	&	0.7	&	 $\boldsymbol{m}_{4, 1}$ 	&	 $\boldsymbol{u}_{4, 1}$ 	&	0.13	&	0.33	&	3.54	&	7.88	&	3.54	&	7.89	&	2.23	&	4.84	&	6.58	\\
		42 	&	150	&	0.7	&	0.7	&	 $\boldsymbol{m}_{4, 1}$ 	&	 $\boldsymbol{u}_{4, 1, r}$ 	&	0.12	&	-0.11	&	3.56	&	6.27	&	3.57	&	6.27	&	1.76	&	3.61	&	4.91	\\
		43 	&	150	&	0.7	&	0.7	&	 $\boldsymbol{m}_{4, 1}$ 	&	 $\boldsymbol{u}_{4, 2}$ 	&	0.12	&	0.66	&	3.56	&	11.40	&	3.56	&	11.42	&	3.20	&	7.20	&	9.79	\\
		44 	&	150	&	0.7	&	0.7	&	 $\boldsymbol{m}_{4, 1}$ 	&	 $\boldsymbol{u}_{4, 2, r}$ 	&	0.13	&	-1.10	&	3.57	&	10.04	&	3.57	&	10.10	&	2.81	&	6.35	&	8.64	\\
		45 	&	150	&	0.9	&	0.9	&	 $\boldsymbol{m}_{4, 1}$ 	&	 $\boldsymbol{u}_{4, 1}$ 	&	0.01	&	0.10	&	0.91	&	5.45	&	0.91	&	5.45	&	5.96	&	6.19	&	5.10	\\
		46 	&	150	&	0.9	&	0.9	&	 $\boldsymbol{m}_{4, 1}$ 	&	 $\boldsymbol{u}_{4, 1, r}$ 	&	0.01	&	-0.05	&	0.91	&	4.19	&	0.91	&	4.19	&	4.58	&	4.77	&	3.93	\\
		47 	&	150	&	0.9	&	0.9	&	 $\boldsymbol{m}_{4, 1}$ 	&	 $\boldsymbol{u}_{4, 2}$ 	&	0.01	&	0.39	&	0.91	&	8.34	&	0.91	&	8.35	&	9.14	&	9.31	&	7.66	\\
		48 	&	150	&	0.9	&	0.9	&	 $\boldsymbol{m}_{4, 1}$ 	&	 $\boldsymbol{u}_{4, 2, r}$ 	&	0.01	&	-0.61	&	0.91	&	7.61	&	0.91	&	7.63	&	8.34	&	8.56	&	7.04	\\
		49 	&	1000	&	0.7	&	0.7	&	 $\boldsymbol{m}_{6, 1}$ 	&	 $\boldsymbol{u}_{6, 2}$ 	&	0.02	&	0.41	&	1.36	&	8.47	&	1.36	&	8.48	&	6.22	&	37.83	&	7.72	\\
		50 	&	1000	&	0.7	&	0.7	&	 $\boldsymbol{m}_{6, 1}$ 	&	 $\boldsymbol{u}_{6, 2, r}$ 	&	0.02	&	-0.08	&	1.36	&	7.83	&	1.36	&	7.83	&	5.75	&	35.09	&	7.16	\\
		51 	&	150	&	0.7	&	0.7	&	 $\boldsymbol{m}_{6, 1}$ 	&	 $\boldsymbol{u}_{6, 2}$ 	&	0.14	&	0.97	&	3.56	&	13.33	&	3.57	&	13.37	&	3.74	&	8.39	&	11.41	\\
		52 	&	150	&	0.7	&	0.7	&	 $\boldsymbol{m}_{6, 1}$ 	&	 $\boldsymbol{u}_{6, 2, r}$ 	&	0.14	&	0.04	&	3.56	&	10.75	&	3.56	&	10.75	&	3.02	&	6.77	&	9.21	\\
		53 	&	1000	&	0.7	&	0.7	&	 $\boldsymbol{m}_{4, 1}$ 	&	 $\boldsymbol{u}_{4, 3}$ 	&	0.02	&	0.04	&	1.36	&	2.72	&	1.36	&	2.72	&	2.00	&	11.30	&	2.31	\\
		54 	&	1000	&	0.7	&	0.7	&	 $\boldsymbol{m}_{4, 1}$ 	&	 $\boldsymbol{u}_{4, 3, r}$ 	&	0.02	&	0.02	&	1.36	&	2.37	&	1.36	&	2.37	&	1.74	&	9.30	&	1.90	\\
		55 	&	150	&	0.7	&	0.7	&	 $\boldsymbol{m}_{4, 1}$ 	&	 $\boldsymbol{u}_{4, 3}$ 	&	0.12	&	0.32	&	3.56	&	5.65	&	3.57	&	5.66	&	1.59	&	3.10	&	4.21	\\
		56 	&	150	&	0.7	&	0.7	&	 $\boldsymbol{m}_{4, 1}$ 	&	 $\boldsymbol{u}_{4, 3, r}$ 	&	0.11	&	0.21	&	3.56	&	4.93	&	3.56	&	4.93	&	1.38	&	2.43	&	3.30	\\
		57 	&	1000	&	0.7	&	0.7	&	 $\boldsymbol{m}_{6, 2}$ 	&	 $\boldsymbol{u}_{6, 4}$ 	&	0.02	&	0.02	&	1.36	&	1.38	&	1.36	&	1.38	&	1.02	&	1.31	&	0.27	\\
		58 	&	150	&	0.7	&	0.7	&	 $\boldsymbol{m}_{6, 2}$ 	&	 $\boldsymbol{u}_{6, 4}$ 	&	0.14	&	0.14	&	3.56	&	3.59	&	3.56	&	3.60	&	1.01	&	0.38	&	0.52	\\
		59 	&	1000	&	0.7	&	0.7	&	 $\boldsymbol{m}_{4, 2}$ 	&	 $\boldsymbol{u}_{4, 4}$ 	&	0.02	&	0.01	&	1.36	&	1.79	&	1.36	&	1.79	&	1.32	&	5.64	&	1.15	\\
		60 	&	150	&	0.7	&	0.7	&	 $\boldsymbol{m}_{4, 2}$ 	&	 $\boldsymbol{u}_{4, 4}$ 	&	0.14	&	0.19	&	3.56	&	4.04	&	3.56	&	4.04	&	1.14	&	1.39	&	1.89	\\	
		\hline
	\end{longtable}
}

\end{document}